\documentclass[12pt]{article}
 \input epsf.sty

\newcommand{\beq}{\begin{equation}}
\newcommand{\eeq}{\end{equation}}
\newcommand{\eqn}{\begin{eqnarray}}
\newcommand{\enn}{\end{eqnarray}}

\def\Label#1{\label{#1}%
  \smash{\hbox to0pt{\raise1ex\hbox{\tiny[#1]}\hss}}}
\def\noLabels{\let\Label=\label}
\def\nobbibitem{\let\bbibitem=\bibitem}

\def\CV{{\cal V}}

\newcommand{\bbibitem}[1]{\bibitem{#1}\marginpar{#1}}

\newcommand{\be}{\begin{equation}}
\newcommand{\ee}{\end{equation}}
\newcommand{\bea}{\begin{eqnarray}}
\newcommand{\eea}{\end{eqnarray}}

\begin{document}

\renewcommand{\thepage}{\arabic{page}}
\setcounter{page}{1}
\noLabels 
\nobbibitem 

\rightline{hep-th/0408054}
\rightline{UPR-1086-T, UNH-04-06}
\vskip 1cm
\centerline{\large \bf Stringy corrections to K\"ahler potentials, SUSY breaking, }
\vskip3mm
\centerline{\large \bf  and  the cosmological constant problem} 
\vskip 1cm

\renewcommand{\thefootnote}{\fnsymbol{footnote}}
\centerline{{\bf Vijay
Balasubramanian${}^{1,}$\footnote{vijay@physics.upenn.edu}
and
Per Berglund${}^{2,}$\footnote{per.berglund@unh.edu},
}}
\vskip .5cm
\centerline{${}^1$\it David Rittenhouse Laboratories, University of
Pennsylvania}
\centerline{\it Philadelphia, PA 19104, U.S.A.}
\vskip .5cm
\centerline{${}^2$\it  Department of Physics, University of New Hampshire,}
\centerline{\it Durham, NH 03824, USA.}

\setcounter{footnote}{0}
\renewcommand{\thefootnote}{\arabic{footnote}}

\begin{abstract}
The moduli of $N=1$ compactifications of IIB string theory can be stabilized by a combination of fluxes (which freeze complex structure moduli  and the dilaton) and nonperturbative superpotentials (which freeze K\"{a}hler moduli), typically leading to supersymmetric AdS vacua.    We show that stringy corrections to the K\"ahler potential qualitatively alter the structure of the effective scalar potential even at large volume, and can give rise to non-supersymmetric vacua including metastable de Sitter spacetimes.   Our results suggest an approach to solving the cosmological constant problem, so that 
the  scale of the 1-loop corrected cosmological constant can be much 
smaller than the scale of supersymmetry breaking.
\end{abstract}

\newpage

\section{Introduction}
\Label{intro}

It has been argued recently that by using a combination of fluxes, D-branes and non-perturbative effects, all the geometric moduli in some string theory compactifications can be stabilized supersymmetrically in a well-controlled region~\cite{kachruI,douglasI}.   Specifically, in certain Type IIB orientifold models, all the complex structure moduli  and the ten-dimensional dilaton are frozen by fluxes and a superpotential generated either by instantons or by gaugino condensation can freeze all the K\"ahler moduli.

Here  we show  that $\alpha'$ corrections to the K\"{a}hler potential  (e.g., \cite{BBHL,grosswitten,otherkahler})  qualitatively change the structure of the effective scalar potential in these theories, even at large volume.    This is initially surprising since one usually neglects such corrections when volumes are large.   However, because the classical scalar potential in these settings has a ``no-scale'' structure (i.e., is independent of the K\"ahler moduli), 
no-scale violating effects in both the K\"ahler potential and the superpotential must be accounted for even at large volume.
Indeed the leading (perturbative) correction to the K\"ahler potential dominates over the leading (non-perturbative) corrections to the superpotential at large volumes,  causing the scalar potential to approach zero from above rather than from below.    Thus, any minimum in the potential at finite volume is protected by a barrier from  the supersymmetric Minkowski vacuum at infinite volume. 

If the flux contribution to the classical superpotential is sufficiently large, it turns out that the K\"ahler moduli cannot be stabilized supersymmetrically.   However, due to stringy
modifications of the metric on the K\"ahler moduli space, a non-supersymmetric minimum can appear in the scalar potential. 
For a range of parameters, this minimum is under reliable semiclassical control, in a region where the well-understood leading perturbative correction to the K\"ahler potential gives an accurate description of the physics. Depending on the sign of the scalar potential at the minimum, this will give rise  to non-supersymmetric AdS, Minkowski and de Sitter vacua.      Thus, the no-go theorems eliminating Kaluza-Klein de Sitter compactifications of M theory \cite{nogo} are avoided by the inclusion of known stringy higher derivative corrections in the effective action~\cite{grosswitten,BBHL,otherkahler}. (For other recent works on how to obtain de Sitter vacua from M-theory, see~\cite{becker,eva1,quevedo2,quevedo1,dealwis,buchbinder}.)

In flat space, once supersymmetry is broken, the gravitino mass, which is the order parameter for supersymmetry breaking, also sets the overall scale for loop contributions to the cosmological constant~\cite{wb}.     This leads to the cosmological constant problem: how can the scale of the cosmological constant be so much smaller than the scale of supersymmetry breaking in the real world?  Our results suggest an approach to this problem -- a non-supersymmetric AdS vacuum can be lifted by one-loop corrections to a de Sitter vacuum with a very small cosmological constant if the supersymmetry breaking scale is of the same order as the AdS scale.   We discuss how these scales are estimated and how this scenario can be realized in our setup.   We also raise the possibility that non-supersymmetric vacua of the kind we describe could tunnel eventually to deeper supersymmetric vacua that lie outside the geometric region of the moduli space.

\section{Stringy corrections to the K\"{a}hler potential}

\subsection{A brief review}

The work of Kachru et al. (KKLT)~\cite{kachruI}, developing from earlier studies including
 \cite{fluxpapers1,fluxpapers2,fluxpapers3,fluxpapers4,ps,gkp}, has shown how all the geometric moduli might be stabilized in  IIB orientifold or F-theory compactifications with $N=1$ supersymmetry in $d=4$ dimensions.  (For a more detailed review, see \cite{reviews}.)  The starting point in the type IIB setting is a string compactification with NS and RR 3-form fluxes, $H$ and $F$ respectively, through the three-cycles of a Calabi-Yau manifold $M$.  (See \cite{fluxpaperstoo} for a sampling of recent work in additional settings.) This gives rise to a classical superpotential~\cite{gvw},
\begin{equation}
W_0
= \int_M \Omega\wedge G\,,\quad G = F - \tau H~,
\Label{e:fluxpotential}
\end{equation}
where $\tau$ is the axion-dilaton. In terms of the effective four-dimensional supergravity theory there is a scalar potential (see, e.g.,~\cite{wb})
\begin{equation}
V=e^K(G^{i\bar j} D_i W D_{\bar j} W-3 |W|^2)~,
\Label{e:sugrapotential}
\end{equation}
where $D_iW=\partial_i W+ W \partial _i K$, $G_{i\bar j}=\partial_i\partial_{\bar j} K$ is the K\"ahler metric derived from the classical  K\"ahler potential,
\begin{equation}
K=- {\rm log}(\int_M \Omega\wedge\bar\Omega) - {\rm log}(\tau-\bar\tau) - 2\, {\rm log}(\int_M J^3)~,
\Label{e:kahlerpotential}
\end{equation}
(with $J$ the K\"ahler form and $\Omega$ the holomorphic $3$-form on $M$)
and $i$ runs over all the scalar fields in the theory, including the complex structure moduli, $z_j$, the dilaton $\tau$, and the complexified K\"ahler moduli, $\rho_k=b_k + i \sigma_k$. Here $b_k$ is an axion charge coming from the RR 4-form and $\sigma_k$ is the volume of a 4-cycle.   We can write
\begin{equation}
\sigma_k = \partial_{t_k} {\cal V} = {1\over 2} \kappa_{ijk} t^i t^j~,
\Label{4to2cycles}
\end{equation}
in terms of $t^i$ which measure areas of 2-cycles and where the classical volume is 
\begin{equation}
{\cal V} = \int_M J^3 = {1\over 6} \kappa_{ijk} t^i t^j t^k~.
\end{equation}
We should understand $\CV$ as an implicit function of the the complexified 4-cycle moduli $\rho_k$ via the relation between $\sigma_k$ and the $t^i$ in~(\ref{4to2cycles}).\footnote{To write the K\"ahler potential in this form a dilaton dependent rescaling of the K\"ahler parameters has been done in both \cite{kachruI} and \cite{BBHL}.  We will discuss the details of this later.}

Supersymmetric vacua are obtained by requiring that $D_i W=0$.   For a generic choice of flux $G$, this extremization of $W$ fixes all the $z_j$ and $\tau$.   Also, since the classical superpotential is independent of the $\rho_k$, 
the $3|W|^2$ cancels out exactly from $V$ 
(see~\cite{gkp} and references therein), leading to a no-scale potential.   The net result, having stabilized the $z_j$ and $\tau$ supersymmetrically, is that the potential  becomes entirely independent of the K\"ahler moduli, which therefore remain free to take any value.   

In order to stabilize the K\"ahler moduli we require an effect that causes the superpotential to depend on $\rho_k$. KKLT proposed two mechanisms~\cite{kachruI}: (I) brane instantons or (II) gaugino condensation. In either scenario, the contribution to the superpotential will be of the general form
 \begin{equation}
W= A_k e^{ i a_k \rho_k}
\Label{e:M-instantonpotential}
\end{equation}
where $A_k$ is a one-loop determinant that depends on the complex structure, $a_k=2\pi/N_k$ and $N_k=1$ for case (I)  while $N_k$ is the dual Coxeter number for the gauge group in case (II). (For more details, see~\cite{kachruI}.)    In fact, there can be an infinite series of such terms from multi-instanton contributions 
\begin{equation}
W \sim \sum_m A_{k,m} e^{i m a_k \rho_k}~.
\Label{multi-instanton}
\end{equation}
We will study moduli stabilization in situations such that $a_k \sigma_k$ is sufficiently larger than $1$ to require only the leading term (\ref{e:M-instantonpotential}).

In both cases  the superpotential can receive a contribution of the form (\ref{e:M-instantonpotential}) from each 4-cycle in $M$ that descends from a divisor in the Calabi-Yau 4-fold with arithmetic genus 1~\cite{witten,katzvafa}. 
(For gaugino condensation to occur the complex structure moduli must also be frozen in such a manner as to admit an enhanced gauge symmetry from coincident 7-branes.  See \cite{shamitgaugino, camara} and references therein for discussions of when this is possible.)  It was shown in \cite{antonella} that for certain Calabi-Yau 4-folds it is possible to choose a basis for the Kahler form $J$ such that {\it every} associated divisor has arithmetic genus 1.  For such F-theory compactifications every K\"ahler modulus will appear in  the non-perturbative superpotential and the scenario of \cite{kachruI} can stabilize all moduli supersymmetrically at finite (and possibly large) volume \cite{douglasI}.\footnote{In fact, it has been recently argued that in the presence of fluxes, the condition on the arithmetic genus can be relaxed, and both the brane instantons and gaugino condensation can occur in a wider class of settings \cite{shamitgaugino}.  See, however, \cite{savbarren} for a contrary argument.}

For a general choice of flux there can be multiple supersymmetric extrema at any of which the contribution to the superpotential from the fluxes is given by~(\ref{e:fluxpotential}).
Thus, from the point of view of the K\"ahler moduli $W_0$ is  constant and  the effective superpotential takes the form
\begin{equation}
W = W_0 + \sum_{k=1}^{h_{11}} A_k e^{ i a_k   \rho_k}~.
\Label{e:superpotentialW0}
\end{equation}
with $h_{11}$ the number of K\"ahler moduli on $M$.
In general the constants $A_k$ and $a_k$ may be different for the different K\"ahler moduli $\rho_k$ depending on which particular mechanism is used to stabilize the moduli.  For purposes of argument we will follow KKLT \cite{kachruI} and take $A_k \sim 1$ to be real, $a_k \sim 2\pi/N$ and $W_0$ to be real and negative.    (The latter turns out to be necessary, given the real $A_k$, for the existence of supersymmetric minima in which the RR 4-forms complexifying the K\"ahler moduli are not turned on.)  Qualitatively, the structure of the potential then looks like Fig.~1A.  Namely, $V \to 0$ from below as the moduli grow large, and there is a supersymmetric AdS minimum for large values of the moduli.

\subsection{Leading no-scale violations in the K\"ahler potential}

Above we discussed the classical no-scale structure of the scalar potential for the K\"ahler moduli and recalled the argument of \cite{kachruI} that non-perturbative effects will produce no-scale violating terms in the superpotential, of which only the leading terms (\ref{e:superpotentialW0}) are relevant at moderate volumes.    In fact, it is  known that the no-scale structure of the classical scalar potential is also violated by stringy 
corrections to the K\"ahler potential~\cite{BBHL}.   
 These   effects are often neglected because they are most important when curvatures are large and thus should be suppressed at large volume.  However, 
 since the classical potential for the K\"ahler moduli vanishes, the leading no-scale violating effects in both the superpotential and the K\"ahler potential are going to be relevant to the qualitative structure of the scalar potential even at moderate and large volumes.  In the present $N=1$ context, there can be perturbative corrections to the K\"ahler potential, worldsheet instantons, and open string effects.  While little is known about these effects in the general $N=1$ setting, the leading perturbative correction, which will be the most relevant term at moderate to large volumes, has been well studied by Becker, Becker, Haack and Louis (BBHL)~\cite{BBHL}.

BBHL, using previous results by other authors including \cite{grosswitten,otherkahler,candelasetal}, 
showed that the leading $\alpha^{\prime 3}$ corrections to the low energy action of the Type II string (which are implied by the underlying special geometry) arise from a dilaton dependent correction to the K\"{a}hler moduli part of the K\"ahler potential.  (Also see the recent paper~\cite{grimmlouis}.)  In the string frame:
\begin{equation}
K= -2 \log\left[ e^{-3\phi_0/2}2 {\cal V}
+ \xi \Big({-i(\tau{-}\bar\tau)\over 2}\Big)^{3/2}  \right] - \log\left[ -i\int_M \Omega \wedge \bar{\Omega} \right] -\log\left[-i(\tau{-}\bar\tau)\right]~.
\Label{eulercorrection}
\end{equation}
Recall that ${\cal V}$ is the classical volume of $M$, where we use~(\ref{4to2cycles}) to express the area of the 2-cycles, $t^k$, in terms of the complexified 4-cycle moduli $\rho_k$.   The factor of $exp(-3\phi_0/2)$ comes from relating the K\"ahler moduli in the Einstein frame to the string frame, and the $\alpha'^3$ correction is given by the term proportional to $\xi= -\zeta(3)\chi(M)/(2(2\pi)^3)$ with $\chi(M)$ the Euler number of $M$.\footnote{We use a slightly different convention than that of~\cite{BBHL} in which geometric objects of the Calabi-Yau are measured in units of $\alpha'$, i.e., we set $\alpha'=1$ rather than $\alpha'=1/(2\pi)$.} Hence the dilaton $\tau$ and the K\"ahler moduli $\rho_k$ are not decoupled from each other in $K$.     When evaluating the K\"ahler potential after stabilization of the dilaton, we can replace $-i(\tau - \bar\tau)/2$ by the fixed value $e^{-\phi_0}$.  However, in order to compute the derivatives required for the connection and the metric on the moduli space we must treat $\tau$ as variable.

BBHL showed that  the $\alpha^{\prime 3}$ correction to $K$ breaks the no-scale structure of the classical supergravity potential (\ref{e:sugrapotential}). In particular, assume that the complex structure moduli $z_i$ and the dilaton $\tau$ have been stabilized by the fluxes at a supersymmetric minimum, such that $D_{z_i} W=D_\tau W=0$, and let the superpotential be given only by the classical contribution from the flux, $W=W_0$.  
The scalar potential then takes the form~\cite{BBHL}
\begin{equation}
V=3 \xi  {(\xi^2 + 7 \xi {\cal V}+ {\cal V}^2)\over ({\cal V}-\xi)(2 {\cal V} + \xi)^2} 
e^K |W_0|^2~. 
\Label{eulerscalar1}
\end{equation}
Thus, even in the absence of the K\"ahler moduli dependent nonperturbative superpotential, $V$ is not a constant -- there is a runaway to large volume where the potential vanishes in the usual manner.   Notice that the potential (naively) diverges at finite volume $\CV = \xi$.   This divergence can be traced to a vanishing of the determinant of the metric, $g_{\rho\bar\rho}$, on the K\"ahler moduli space at $\CV = \xi$ in this approximation.     However, at such small volumes additional corrections will be important.   
In Sec.~3 we will argue that the structure of the metric on the moduli space can lead to supersymmetry breaking minima in the scalar potential that can be reliably understood using only the  leading corrections to the K\"ahler potential (\ref{eulercorrection}) because they 
are located at sufficiently large volume.

With this in mind we turn to combining the  leading correction to the K\"ahler potential with the leading term in the non-perturbative superpotential (\ref{e:M-instantonpotential}).   In their original work, KKLT  did not include $\alpha'$ corrections to the K\"ahler potential as the K\"ahler moduli could be stabilized supersymmetrically at large values where the corrections do not significantly affect the structure of the vacuum~\cite{kachruI}.\footnote{This has also been confirmed independently by Berg, Haack and Kors~\cite{berg}.}  However, because of their nature as perturbative rather than non-perturbative effects, the $\alpha^{\prime 3}$ corrections give the dominant behavior for large volume. In particular,  the scalar potential has to have a maximum at which $V>0$ (given $\xi>0$) after any moduli stabilizing (possibly supersymmetric) minimum.\footnote{The condition on $\xi$ is equivalent to requiring that $\chi(M)<0$, which occurs generically for manifolds, $M$, with a small number  of K\"ahler moduli.}

 To see this, consider the scalar potential, once again assuming that the complex structure moduli and the dilaton satisfy $D_{z_i} W=D_\tau W=0$, respectively.    The scalar potential $V$ takes the form
\begin{eqnarray}
V
&=&e^K\Big[ G^{\rho_j\bar\rho_k}\Big(a_j A_j \bar a_k \bar A_k e^{a_j \rho_j +\bar a_k \bar \rho_k}
 +
 a_j A_j e^{a_j \rho_j} \bar W \partial_{\bar \rho_k} K +
\bar a_k \bar A_k e^{\bar a_j \bar \rho_j} W \partial_{\rho_j} K\Big) \cr
&+& 3 \xi  {(\xi^2 + 7 \xi {\cal V}+ {\cal V}^2)\over ({\cal V}-\xi)(2 {\cal V} + \xi)^2} |W|^2\Big]~.
\Label{eulerscalarpotential}
\end{eqnarray}
While the first term is positive, it is readily shown by taking the requisite derivatives that the second and third terms are negative when the $\rho_j$ are large.   
Thus, without the $\alpha'$ correction, that is $\xi=0$, the scalar potential is dominated by the second and third term  and would approach $V=0$ from below for large volume. However, with the $\alpha'$ corrections included the fourth term is the leading one and so the potential approaches $V=0$ from above (see Fig.~1B).    Note that the potential still diverges at $\CV = \xi$.

Provided we are able to stabilize the K\"ahler moduli at sufficiently large volume,  (\ref{eulerscalarpotential}) will provide a reliable description of the scalar potential in the vicinity of the vacuum and at larger volumes.  As we have noted, at small volumes there are further important corrections to the K\"ahler potential about which little is known in the $N=1$ setting.   Since we will argue in Sec.~3 that the existence of a singularity in (\ref{eulerscalarpotential}) at the small volume $\CV = \xi$ already has important effects in the large volume region where the potential is reliable, it is important to get some further understanding of the origin and nature of the singularity and how it might be affected by the additional corrections.

\begin{figure}
\Label{fig:potential1}
  \begin{center}
 \epsfysize=2in
   \mbox{\epsfbox{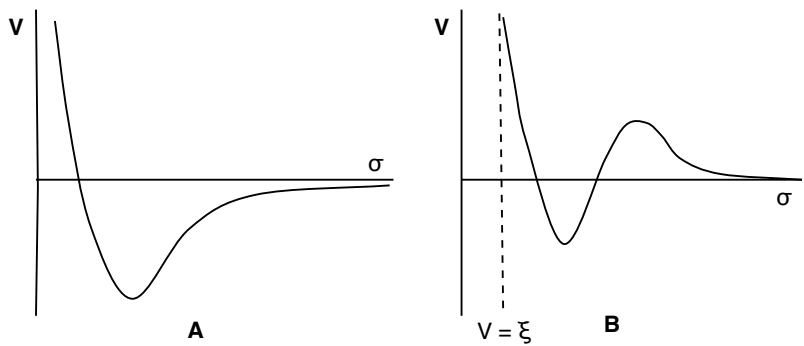}}
    \caption{General form of scalar potential for K\"{a}hler moduli.  For convenience, only one K\"{a}hler parameter is shown.  (A) Potential with non-perturbative superpotential generated by 3-brane instantons or gaugino condensation.   (B) Effects of the $\alpha'^3$ correction to the K\"{a}hler potential.  The scalar potential approaches zero from above.   The situation depicted is for sufficiently small $W_0$ -- a supersymmetric AdS minimum persists.  The dashed line is the location of the (naive) singularity in the potential at $V = \xi$. }
 \end{center}
\end{figure}

\subsection{Insights from $N=2$ supersymmetry}

In the $N=2$ context of type IIB theory compactified on a Calabi-Yau manifold $M$ the 
moduli space for the hypermultiplets is a $4(h_{11}+1)$ (real) dimensional hyperk\"ahler manifold. Since the dilaton is in a hypermultiplet, quantum corrections to the hypermultiplet moduli space are possible.
 We are interested in studying the restriction of this moduli space spanned by the $h_{11}$  complexified K\"ahler moduli $x^i$ whose imaginary parts survive the orientifolding to $N=1$ SUSY.
Using mirror symmetry and special geometry, the world-sheet instanton corrected $N=2$ K\"ahler potential for the 
$x^i$, 
(with $t^i={\rm Im}(x^i)$ the area of a 2-cycle) is given by~\cite{candelasetal}
\eqn
K_{N=2}(x^k,\bar x^k) &=& - {\rm log}\Big[{i\over 6} \kappa_{ijk} (x^i{-}\bar x^i)(x^j-\bar x^j)(x^k-\bar x^k)+4 \xi +\cr
&&
\sum_{\{n_k\}} \tilde N_{\{n_k\}} (e^{2\pi i n_kx^i}+e^{-2\pi i n_k\bar x^i})(1-i\pi n_k(x^i-\bar x^i))\Big]~.
\Label{N=2Kahlerpotential}
\enn
where $\kappa_{ijk}$ and $\xi$ are given as in the $N=1$ case, 
and the $\tilde N_{\{n_k\}}$ are closely related to the instanton numbers of degree $\{n_k\}$. 
Note that  at the perturbative level (second term) the $N=2$ K\"ahler potential only receives corrections proportional to $\alpha'^3$ (which we have set to 1)
while the last term corresponds to non-perturbative world-sheet instanton corrections.

Given the above $N=2$ K\"ahler potential we can perform a naive reduction to $N=1$ supersymmetry along the lines of~\cite{dauriaetal}. The $N=1$ K\"ahler potential is given by $\hat K(\rho_k,\bar \rho_k)=2 K_{N=2}(x^k,\bar x^k)$ where we express the 2-cycle moduli $x^k$ on the RHS in terms of the 4-cycle moduli $\rho_k$.\footnote{Strictly speaking, we have to rescale the $x^k$ as well as the $\rho_k$ by $exp(\phi/2)$ and  $exp(\phi)$, respectively when relating string frame to Einstein frame. But since we assume that the dilaton has been stabilized by the flux, we can neglect the effect of this rescaling.}
To further simplify the situation we include neither a non-perturbative superpotential nor the worldsheet instanton corrections to the K\"ahler potential. The scalar potential then takes the form
\begin{equation}
\hat V =3   {\xi\over 4 {\cal V}-\xi} 
e^{\hat{K}} |W_0|^2~.
\Label{decoupling}
\end{equation}
To leading order in $\xi/\CV$, when $\CV>>\xi$, 
$\hat V$ agrees with the scalar potential (\ref{eulerscalar1}) obtained using the correct $N=1$ analysis.   
At small values of the moduli, and hence small volume, both scalar potentials diverge. However, the singularity appears for somewhat different values of $\CV$.\footnote{Interestingly, one can show that when including the $\alpha^{'3}$ corrections, the determinant of metric derived for the moduli space of two-cycles, $g_{x^i\bar x^i}=\partial_{x^i}\partial_{\bar x^i} (K_{N=2})$ has the same singular behavior as the determinant of the metric derived for the $N=1$ K\"ahler potential, after including the coupling to the dilaton in the latter.} This is not too surprising since this is a region where higher order corrections to the K\"ahler potential will become important. 
Thus, 
the naive $N=1$ reduction of the $N=2$ K\"ahler potential does not precisely reproduce all the features of the BBHL analysis of the $N=1$ K\"ahler potential but has the same qualitative features.    It is of interest to consider the behavior of the metric on the moduli space when the non-perturbative $N=2$ corrections are taken into account, as this may give an indication of what to expect when large quantum effects are present. In particular, it may tell us about the origin of the singular scalar potential.

To be explicit, let us consider the metric on the (complexified) K\"ahler moduli space for the Calabi-Yau manifold, $M_{P^4[5]}$, given by a quintic hypersurface in $P^4$~\cite{candelasetal}.
In terms of the complex parameter $\psi$ which describes the one complex structure deformation of the mirror manifold, $\tilde M$, this metric has a logarithmic singularity, $g_{\psi\bar \psi}\sim -{\rm log}(|\psi|-1)$ as $\psi\to 1$, the conifold point.   Thus, in terms of the  coordinate $x$ relevant for the K\"ahler deformation of $M$, the metric including all worldsheet instanton corrections is
\begin{equation}
g_{x\bar x} = \Big|{{\rm d} \psi\over {\rm d} x}\Big|^2 g_{\psi\bar \psi} \sim -{1\over {\rm log}(\psi-1)}\quad{\rm as}\quad \psi\to 1~,
\end{equation}
where we have used that ${\rm d} \psi/ {\rm d} x\sim -({\rm log}(|\psi|-1))^{-1}$~\cite{candelasetal}. Thus, $g_{x\bar x}$ vanishes as we approach the location which in $\tilde{M}$ is the conifold point.\footnote{From mirror symmetry we know that the complex structure moduli space of $\tilde{M}$
is isomorphic to the K\"ahler moduli space of $M$, and vice versa with $\tilde{M}$ and $M$ interchanged. Thus, to reduce verbiage we will also loosely refer to the point $x(\psi=1)$ in $M_{P^4[5]}$ as the conifold point.}  Furthermore, using the  mirror map between $\psi$ and $x$ one can show that $\psi= 1$ corresponds to $x\sim i 1.2$~\cite{candelasetal}. Thus, the size of the 2-cycle is finite at the the conifold point.  However, in the reduction from $N=2$ to $N=1$ supersymmetry the relevant K\"ahler modulus is the complexified four-cycle modulus, $\rho$. Just as above one can show that
\begin{equation}
g_{\rho\bar \rho} = \Big|{{\rm d} \psi\over {\rm d} \rho}\Big|^2 g_{\psi\bar \psi} \sim -{1\over {\rm log}(\psi-1)}\quad{\rm as}\quad \psi\to 1~,
\end{equation}
where the important part of ${\rm d} \psi/ {\rm d} \rho$ as $\psi \to 1$ again goes as $-({\rm log}(|\psi|-1))^{-1}$ as follows from~\cite{candelasetal}. Furthermore, $\rho$ is non-zero at $\psi=1$ and hence the size of the 4-cycle is also finite at the conifold.

Thus, after including the worldsheet instanton corrections, the singularity in the scalar potential for the K\"ahler moduli of IIB string on $M$ is associated with a conifold singularity in IIA string theory on $\tilde{M}$.  One might wonder if quantum effects resolve this singularity.   In fact, Ooguri and Vafa showed that the fully quantum corrected $N=2$ hyperk\"ahler metric obtained in type IIA on $\tilde M$ is not singular at the conifold locus~\cite{ov}. This is due to Euclidean D2-branes wrapping the shrinking three-cycle at the conifold of $\tilde{M}$.   In type IIB on $M$ this phenomenon is translated into D5-branes which at the mirror of the conifold locus wraps a shrinking six-cycle~\cite{ps,gk}. Still, while the quantum corrected metric $g_{\psi\bar\psi}$ is finite it follows from the naive mirror map between $\psi$ and $\rho$ that $g_{\rho\bar \rho}$ vanishes at the conifold.    It may be that the mirror map also  receives quantum corrections that modify the behavior of $g_{\rho\bar{\rho}}$ at the conifold locus.
In any case,  the naive $N=1$ reduction of the $N=2$ K\"ahler potential for $M_{P^4[5]}$ indicates that the divergence in the scalar potential occurs because $g^{\rho\bar\rho}\to\infty$ at the conifold point.    A full $N=1$ analysis should be performed to see whether this explanation survives the inclusion of all of the higher order $N=1$ corrections.

It is interesting to note that the $N=1$ reduction of the worldsheet instanton corrected $N=2$ analysis reproduces features of the $N=1$ potential (discussed in Sec.~2.2) that used only the leading perturbative correction: (a)  there is a singularity in the potential at finite volume, (b) the (determinant of) the metric on the moduli space declines as we approach this locus. 
In Sec.~2.2 it was not clear why $V$ diverges except that the determinant of the metric vanishes. The above naive $N=1$ reduction of the $N=2$ analysis indicates the apparent relationship between the divergence of the scalar potential and a conifold degeneration in the mirror manifold.

  Given the example of the $M_{P^4[5]}$ above we should expect such behaviour to be universal in the vicinity of conifolds even in examples with many K\"ahler moduli.     We also expect that the higher order $N=1$ corrections will not change the qualitative  features we are describing here.   We have already seen in Sec.~2.2  that the leading perturbative term preserves these features, and in the next section we will find minima of the scalar potential in regions far from the singularity where this leading term dominates over the higher order perturbative and non-perturbative corrections that are possible.

\section{Supersymmetry breaking}
\Label{SUSYbreaking}

Several scenarios have been proposed for completely breaking supersymmetry to achieve a metastable de Sitter minimum in  KKLT-like settings.    In their original work KKLT proposed adding an anti-brane to their compactification breaking supersymmetry and contributing a positive power law to the scalar potential \cite{kachruI}.   The resulting potential had a de Sitter minimum and  approached zero from above at infinity.    Silverstein and Saltman pointed out that stabilizing the complex structure moduli at a non-supersymmetric point (so the $D_i W \neq 0$ for some of them) would have a similar effect -- the $e^K (G^{ij}D_iW D_{\bar{j}}W)$ term in the supergravity potential (\ref{e:sugrapotential}) produces a power law that can lift the KKLT AdS minimum to a de Sitter minimum \cite{eva1}.    Burgess et.al. proposed that adding fluxes on the D7-branes in the compactification would achieve a similar effect \cite{quevedo1}.    Further possibilities were pointed out by Escoda et al~\cite{quevedo2}, Brustein and de Alwis~\cite{dealwis}, Becker et al.~\cite{becker} and by Buchbinder~\cite{buchbinder}.   Here we observe that when the tree-level superpotential $|W_0|$ is sufficiently large compared to the non-perturbative contribution, supersymmetry will be broken.    Furthermore, in this context, the existence of conifold loci
 modifies the metric on the K\"ahler moduli space in such a manner as to admit non-supersymmetric AdS, Minkowski and metastable de Sitter soutions.

\subsection{Large $|W_0|$ and supersymmetry breaking}
    Given that we have already stabilized the complex structure moduli and the dilaton supersymmetrically using fluxes, supersymmetric minima exist when $D_{\rho_k} W = 0$ for all K\"ahler moduli.  The nonperturbative superpotential is of the form
    \begin{equation}
    W = W_0 + W_{{\rm np}} = W_0 + \sum_{j=1}^{h_{11}} A_j \,  e^{i a_j \rho_j}
    \Label{thesuperpotential}
    \end{equation}
Here we have assumed that volumes are sufficiently large to ignore multi-instanton terms in the superpotential.
 The supersymmetry condition $D_{\rho_k} W = 0$ is solved as:
\begin{equation}
W_0 = -  \left[\sum_{j=1}^{h_{11}} A_j \, e^{ia_j \rho_j} + { i a_k A_k e^{ia_k \rho_k}\over \partial_{\rho_k} K} \right]
\Label{SUSYconditiona}
\end{equation}
In order to find a solution the phases on the left and right hand sides of the equation must be appropriately matched.   For purposes of argument let us simply assume here that all the $A_k$ are real and that $W_0$ is real (a more general setting is easily handled). Generically, the $\rho_k$ will not be purely imaginary, and $Re(\rho_k)$ are determined such that $W$ is real.

\paragraph{Small $|W_0|$ and supersymmetric vacua: } First we show that for any sufficiently small $|W_0|$ there will be a supersymmetric AdS minimum at large volume.  At large volume it is sufficient to keep the leading (nonperturbative) corrections to the superpotential and the leading (perturbative) correction to the K\"ahler potential (\ref{eulercorrection}).      The supersymmetry condition is then of the form
\begin{equation}
W_0 = -  \left[\sum_{j=1}^{h_{11}} A_j \, e^{ia_j \rho_j} + 2 a_k A_k e^{ia_k \rho_k}{ ( {\cal V} + {\xi\over 2}) \over  t^k} \right]
\Label{eulersusy}
\end{equation}
where we have used that
\begin{equation}
\partial_{\rho_k} K
={i t^k\over 2 {\cal V}+ \xi}~.
\end{equation}
 Since the absolute value of the RHS of (\ref{eulersusy}) is 
 decreasing for sufficiently large $\rho_k$, and there are $h_{11}$ (complex) equations in the $h_{11}$ (complex) unknown values of $\rho_k$,
 it is clear that for any sufficiently small choice of $|W_0|$ there will generically be a discrete set of supersymmetric minima of the scalar potential.    In \cite{kachruI} it was argued that the requisite small values of  $W_0$ can be obtained by tuning fluxes and this argument has been placed on firm footing by the work of \cite{ashokdouglas, douglasdenef, giryavets}.

\paragraph{Large $|W_0|$ and supersymmetry-breaking minima: }   We will now argue that for sufficiently large $|W_0|$ there is no supersymmetric minimum in the geometric region\footnote{By geometric region we refer to the part of moduli space for which an  expansion in $\alpha'$ is valid.} of the moduli space, but that new non-supersymmetric minima appear.   A clean analysis of this would require us to include all the higher perturbative and non-perturbative corrections to the K\"ahler potential, $K$, and the superpotenial.  However, as we have discussed, only the leading perturbative correction to $K$ is well understood in the $N=1$ setting.  Thus, the general strategy is to study properties of the leading approximation to the scalar potential as a function of $|W_0|$, but to only trust features that appear at sufficiently large volume where this functional form is a good approximation to the full potential.

\begin{figure}
\Label{fig:potential2}
  \begin{center}
 \epsfysize=3in
   \mbox{\epsfbox{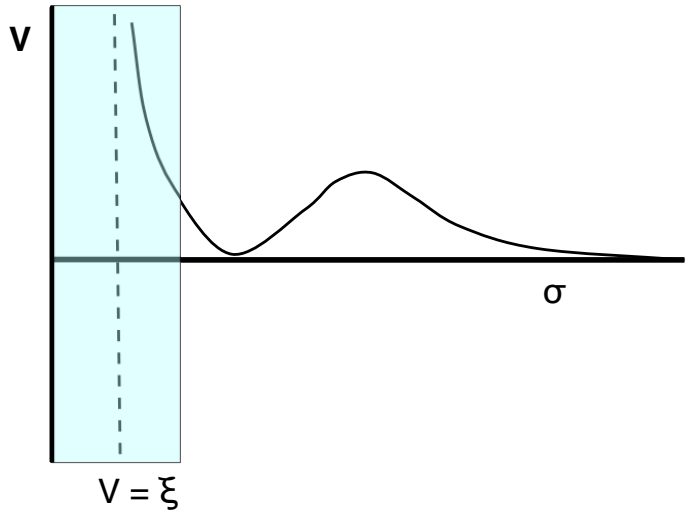}}
    \caption{Non-supersymmetric Minkowski vacuum when the flux contribution to the superpotential, $W_0$, is large.  An accurate description of the shaded region requires inclusion of additional corrections, but for a range of parameters the minimum lies in a region where the leading terms in the scalar potential, $V$, are sufficient.  The dashed line is the location of the (naive) singularity in $V$ at $V = \xi$, after including $\alpha'^3$ corrections to the K\"ahler potential.}
 \end{center}
\end{figure}

Recall that the scalar potential (\ref{eulerscalarpotential}) has a singularity at $\CV = \xi$ which we interpret as saying that the geometric region of the moduli space lies at larger volumes $\CV > \xi$.    
The condition $\CV > \xi$ then cuts off the apex of the classical K\"ahler cone.   Now,  for any given $|W_0|$, a solution to  the supersymmetry condition (\ref{eulersusy}) determines all the K\"ahler moduli.   There are $h_{11}$ equations in $h_{11}$ unknowns generically giving rise to  a discrete set of solutions.    Consider a choice of $|W_0|$ that leads to a solution of (\ref{eulersusy}) on a surface of some fixed classical volume $\CV > \xi$.   Then the $t^k$ in this solution must be bounded away from zero to achieve supersymmetry, because if $t^i = 0$ for some $i$, then $|W_0| = \infty$ by (\ref{eulersusy}) and thus all the $t^k = 0$ leading to vanishing volume for the Calabi-Yau which is counterfactual in the allowed region $\CV > \xi$.    Turning this around, if we seek a supersymmetric minimum in the allowed region $\CV > \xi$, there is an upper bound on $|W_0|$,
\begin{equation}
|W_0| \leq W_{{\rm max}} \, .
\end{equation}
This argument  does not guarantee that for every $\CV > \xi$ there is a choice of $|W_0|$ yielding supersymmetric minima.    It simply shows that there is some upper bound on $|W_0|$ leading to supersymmetric minima in the geometric region of the moduli space.  Any minima that we find in this region for $|W_0| > W_{{\rm max}}$ will break supersymmetry.

 One can see that the larger the value of the $|W_0|$ on the LHS of the supersymmetry condition (\ref{eulersusy}), the smaller the required values of $t_k$ on the RHS and thus the smaller the overall volume.       If we track the line of minima in the potential that are obtained as we increase $|W_0|$ from a very small value that leads to a supersymmetric AdS vacuum,  there are two possibilities: (a) supersymmetric AdS minima ($D_iW = 0$ while $W \neq 0$) could persist all the way to $\CV = \xi$ or (b) supersymmetry breaking could occur at some $\CV > \xi$. Let us consider possibility (b) first.   In this case, when $|W_0| = W_{{\rm max}} - \epsilon$ there will be a supersymmetric AdS minimum with $V = -3 e^K |W|^2$.   Increasing $|W_0|$ to $W_{{\rm max}} + \epsilon$ will break supersymmetry, but since the potential is a continuous function of the $|W_0|$ there will still be an AdS minimum.   For  sufficiently large $|W_0|$, the non-perturbative correction to the superpotential becomes negligible compared to the classical contribution, and thus the potential reduces to the form (\ref{eulerscalar1}), monotonically decreasing from infinity at $\CV = \xi$ to 0 at $\CV = \infty$.    This could happen in one of two ways.  The first possibility is that as $|W_0|$ is increased, the minimum rises to a Minkoswki and then de Sitter minimum and then finally disappears.   If this happens metastable de Sitter space is realized because of the barrier in the potential that we have already described.    The other possibility is that the AdS minimum approaches the $\CV = \xi$ locus  as $|W_0|$ is increased.
  In either this situation, or in case (a) where the supersymmetric minima persist all the way to $\CV = \xi$, there is some critical value $|W_0| = W_{{\rm crit}}$ where an AdS minimum crosses the $\CV = \xi$ locus.\footnote{One can easily show that case (a) always arises if there is only one K\"ahler modulus.}

 In these latter cases, when $|W_0| = W_{{\rm crit}} - \epsilon$, we known that there is an AdS minimum in the geometric region $\CV > \xi$.     Now observe that the scalar potential is a manifestly continuous function of $W_0$ almost everywhere -- the only subtlety can be at the locus $\CV = \xi$,   
where $V$ will diverge for generic values of $W_0$.  (This is because the terms in the last three lines of (\ref{eulerscalarpotential}) all diverge at $\CV = \xi$, and all have different dependences on $W$.)  So if we increase $|W_0|$ to $W_{{\rm crit}} + \epsilon$,  by continuity the scalar potential will still be negative in the vicinity of the former minimum.    Since we know that the potential 
diverges at $\CV = \xi$, while going to zero from above at large volume in all directions, there must now be a non-supersymmetric AdS minimum in the geometric region.   Flat directions will generically not occur since there are as many equations, $\partial_{\sigma_k} V=0$, as unknowns, $\sigma_k$, and because we only expect isolated minima in the context of  $N=1$ supersymmetry.   Finally,   for sufficiently large $|W_0|$,  as discussed above, the potential will be monotonically decreasing from infinity at $\CV = \xi$ to 0 at $\CV = \infty$.  Thus we see that as $|W_0|$ is increased from the critical value,  non-supersymmetric AdS, Minkowski, and de Sitter vacua will occur until eventually the local minimum disappears.   (See Fig.~2.)

The arguments we made were based on very general considerations of the qualitative structure of the scalar potential and supersymmetry conditions.   In Sec.~2.3 we argued that naive $N=1$ reduction of the well-understood $N=2$ analysis of conifold singularities  reproduced the essential qualitative features of the BBHL $N=1$ potential   that we have explored in detail above.  Since our $N=2$ discussion included stringy and quantum corrections, there is some reason to hope that the necessary qualitative structure is present even at small volume in the $N=1$ setting,  where higher order perturbative and non-perturbative corrections are important, at least when the $N=1$ theory is obtained by orientifolding an $N=2$ compactification.
If so, the same continuity arguments given above will lead to non-supersymmetric minima in the scalar potential at small volume, and AdS, Minkowski and metastable de Sitter minima will all occur.

\paragraph{Toy model:}
Our analysis was carried out keeping only the leading terms in the K\"ahler potential and superpotential.  The question now becomes whether by tuning of $W_0$ such non-supersymmetric minima can be pushed to the region in which this approximation is justified.  To see that this is indeed the case, we can study a toy example along the lines pursued in KKLT \cite{kachruI}.  Consider a situation with one K\"ahler parameter, and a superpotential (\ref{thesuperpotential}) with $A=1$ and $a = 2\pi/10$.   Let us also take the $\alpha'^3$ correction $\xi/2 = \zeta(3) \chi / 4(2\pi)^3 = 0.2$ as it would be for a Calabi-Yau manifold with an Euler number of the order of $\chi = -200$ (the quintic example in Sec.~2.2, for example).  Then for $W_0 \sim -1.7$ there exists a non-supersymmetric Minkowski minimum in the potential where the K\"ahler modulus is stabilized at $\sigma\sim 5$ which gives a classical volume $\CV \sim 2$. Note that while this is a relatively small value, still the $\alpha'^3$ correction is only in the order of 10\% since $\CV/(\xi/2)\sim 10$. Furthermore, $a\, \sigma \sim 3$.  
The higher order perturbative and non-perturbative corrections  are suppressed at such volumes.  
The above toy model shows that it is indeed possible to find non-supersymmetric solutions in a region where the volume of $M$ is (relatively) large such that the higher order corrections can be ignored.

\paragraph{An upper bound on $|W_0|$ and and the discretuum: }
Much of the recent work on distributions of flux vacua \cite{ashokdouglas,douglasdenef,giryavets} has considered whether one can achieve the very small values of $|W_0|$ necessary for supersymmetric vacua at large  volume.  Our considerations require rather that sufficiently large values of $|W_0|$ can be attained.   In the absence of D3-branes, as in our scenarios, a constraint on the values of $|W_0|$ arises because  there is a tadpole cancelation condition of the form
\begin{equation}
{\chi(X) \over 24} =  {1 \over 2 \kappa_{10}^2 T_3} \int_M F \wedge G
\Label{tadpole}
\end{equation}
where $\chi(X)$ is the  Euler character of the Calabi-Yau   4-fold, $X$, on which F-theory is compactified. Equivalently, by taking  the orientifold limit it is the net D3-charge from O3-planes and induced charges on  7-branes in the IIB picture.  The RHS of (\ref{tadpole}) involves a wedge product between the $H$ and $F$ fluxes, while the flux superpotential $W_0 = \int_M \Omega \wedge (F -\tau H)$ arises from a wedge product between these fluxes and the holomorphic 3-form.  Because the products are different, (\ref{tadpole}) does not immediately translate into a constraint on $|W_0|$.  However, all else being equal, we can estimate that in order to achieve the tadpole cancelation condition the components of $H$ and  $F$ that are orthogonal to each other will scale as $|H| \sim |F| \sim \sqrt{\chi}$.   Here we are leaving out factors of $\alpha^\prime$ and $g_s$ as well as numerical factors.   Note that there is no constraint on the components of $H$ and $F$ that are parallel to each other and these can also contribute to the superpotential.    Even if we ignore such parallel components of $H$ and $F$, in view of the above we estimate that the maximum value that $|W_0|$ can achieve will at least scale as $\sqrt{\chi}$.    Since the 4-folds giving rise to IIB models with a small number of K\"ahler moduli typically have Euler characteristics in the 1000s, there should be no difficulty in achieving $|W_0|$ large enough to break supersymmetry in the manner discussed above.

The values that $W_0$ can take in this analysis form a discrete set.  However, following the arguments of Bousso and Polchinski \cite{bousso} and \cite{ashokdouglas,douglasdenef,giryavets}, the actual values that $W_0$ assumes will be very closely spaced, so much so that they will form a ``discretuum''.   Indeed, for all practical purposes we can think of $W_0$ as a continuously varying parameter.   Thus in our setting $|W_0| - W_{{\rm max}}$ is a supersymmetry breaking parameter that varies essentially continuously as the fluxes are tuned.

\vspace{5mm}
   It is also interesting to note that the new non-supersymmetric minima that we discuss 
are closely related to the  locus in the K\"ahler moduli space which is related to the conifold in the mirror manifold.
   It has been argued in \cite{douglasdenef,giryavets} that when fluxes freeze complex structure moduli, the resulting stable vacua tend to cluster near conifold points. (For another argument in favor of enhanced symmetry points, see \cite{kofmanetal}.)  It is tempting to speculate, particularly in view of mirror symmetry, that similar concentrations of vacua may occur in the K\"ahler moduli space partly via the mechanisms described in this paper.

\subsection{On the cosmological constant problem}
Once supersymmetry is broken, the lack of Bose-Fermi degeneracy means that there will be a loop corrections to the cosmological constant.\footnote{We thank Shanta de Alwis and Bobby Acharya for discussions related to this section.}   In flat space, the supersymmetry breaking scale is related to the gravitino mass $m_\lambda$ \cite{wb} and the one-loop cosmological constant will be proportional to $m_\lambda^2 M_P^2$ where $M_P$ is a cutoff on the effective field theory.   The precise coefficient is model dependent because, for example, the contribution to $\Lambda$ from a particular superfield could be positive or negative (see, e.g., \cite{gaillard}) depending on whether bosons or the fermions have been lifted.   But since all mass splittings will be proportional to $m_\lambda$ (see \cite{wb} for example), we will always find that 
\begin{equation}
\Lambda_{{\rm 1-loop}} = C_{{\rm model}} \, m_\lambda^2 M_P^2\,
\Label{1loopcorrect1}
\end{equation}
where $C_{{\rm model}}$ is model dependent.   (Higher loops will be suppressed by powers of the effective coupling.) This is one precise version of the cosmological constant problem: once supersymmetry is broken, the one-loop contribution to the cosmological constant is set by the supersymmetry breaking scale, a theoretical prediction that is badly violated by the real world.   Anthropic considerations cannot solve this problem unless it is shown that there actually exists some mechanism to stabilize the cosmological constant at a scale much smaller than the supersymmetry breaking scale.  Indeed, even the KKLT scenario \cite{kachruI} suffers from this -- although the cosmological constant is tuned to be small and positive, by adding anti-branes to an AdS vacuum, the large supersymmetry breaking scale from anti-branes will be communicated by interactions into a large one-loop correction to the bulk cosmological constant.

If the cosmological constant arising from the supergravity potential with stringy and instanton corrections does not vanish, further care is necessary.  This is because in the presence of a cosmological constant, the gravitino mass is never zero even if  $\Lambda < 0$ and supersymmetry is preserved.   Indeed, in an AdS background the gravitino mass is \cite{wb} 
\begin{equation}
m_{\lambda} = e^{K/2} |W|
\Label{susygravitino}
\end{equation}
even if the vacuum is supersymmetric. Since the graviton remains massless, we see that Bose-Fermi splittings need not vanish in a supersymmetric AdS vacuum.  Nevertheless loop corrections will vanish because of the properties of propagators in AdS space and the value of the scalar potential will be fixed at $V = -3 m_\lambda^2$.     If supersymmetry is broken, some variations of the superpotential, $D_\alpha W$, will not vanish, and this relation between $V$ and $m_\lambda$ will fail even before loops are accounted for.  The difference will provide an order parameter for supersymmetry breaking.

In view of all this, our results suggest a simple route around the cosmological constant problem.  If we start in a non-supersymmetric AdS vacuum, and tune the breaking scale appropriately, the one-loop corrected cosmological constant
\begin{equation}
\Lambda=\Lambda_{\rm{AdS}}+\Lambda_{\rm{1-loop}}
\end{equation}
 can have a scale much smaller than the supersymmetry breaking scale.   The scenarios considered in this paper naturally realize this possibility.  As we have discussed, as the tree level superpotential is raised from its critical value, there is a line of non-supersymmetric AdS vacua whose vacuum energy decreases to 0 as $|W_0|$ increases.  At the same time, as $|W_0|$ increases, the supersymmetry breaking scale increases from 0 at $|W_0| < W_{{\rm max}}$ to larger positive values.   Thus, assuming a positive one-loop contribution to the vacuum energy, there will necessarily be values of $|W_0|$ at which the one-loop correction will be comparable to the bare vacuum potential energy, thus leading to a net positive cosmological constant close to zero.    

A key ingredient in this argument is once again the fact that the possible values of $W_0$ form a ``discretuum'' \cite{bousso,ashokdouglas,douglasdenef,giryavets} and thus can be tuned with high precision.  Given this argument establishing the plausible existence of vacua with a small cosmological constant despite large mass splittings between superpartners, anthropic considerations can serve as a selection principle.      In fact, this sort of idea had  been  suggested some time ago by  Nilles~\cite{nilles},\footnote{We thank Brent Nelson for pointing out this reference.} and we are arguing that it will be naturally realized here.  The basic idea is applicable to any string compactification setting in which the supersymmetry breaking scale in an AdS minimum can be tuned to be comparable to the AdS scale.

While we will not attempt to review the many other recent approaches to tuning the cosmological constant in both stringy and brane world settings, it is worth comparing our remarks to those of Bousso and Polchinski \cite{bousso}. They noted that the many fluxes present in M-theory can allow a classical and anthropic \cite{anthropic} tuning of the cosmological constant to a small value  by adding a positive energy density to a bare negative cosmological constant.   We are pointing out that in our setting the scale of supersymmetry breaking is set by fluxes and thus can be discretely tuned so that the net, one-loop corrected, cosmological constant can be much smaller than the scale of Bose-Fermi splitting.  (See \cite{susyscale} for other comments on the supersymmetry breaking scale in the context of the landscape of string vacua, and \cite{evascalars} for  related ideas for fine tuning scalar masses.)

\section{Dynamical transitions?} 
In the previous section we explained that if the tree level superpotential is sufficiently large, a supersymmetric minimum of the scalar potential cannot be found in the geometric region of the moduli space.   However, it could well be that there is a supersymmetric extremum in a non-geometric phase that can lie beyond the apparent singularity in the scalar potential.      One reason why it is important to consider such scenarios is that dynamical transitions might be possible between between the non-supersymmetric vacua we discuss and supersymmetric vacua with stabilized moduli that lie in a non-geometric region of the moduli space.   

In order to understand whether this kind of transition can occur, besides determining which of the (meta)stable vacua has lowest energy, one needs to understand whether instantons exist at all for tunneling between spacetimes with different cosmological constants.\footnote{We thank Tom Banks for useful exchanges on this subject.}  The classic work of Coleman and de Luccia (CdL) showed that transitions between universes with vanishing and negative cosmological constants can only occur if the difference between their vacuum energies exceeds a certain bound \cite{CDL}.  What is more, even if such a transition occurs, they argued that the result is not a stable AdS space, but rather an FRW universe with a Big Crunch.       Banks \cite{banks} has argued, following \cite{weinberg} and \cite{mirjam}, that transitions between supersymmetric AdS vacua are not possible because of a CdL-like bound, that transitions between non-supersymmetric AdS vacua would likely lead to singular collapse, and that there are questions about the consistency of such transitions in light of the AdS/CFT correspondence.   But the question of possible decays from and to asymptotically locally AdS spacetimes is not yet  settled\footnote{For example, some asymptotically locally AdS backgrounds can decay via bubbles of nothing \cite{BON}.}  and we should explore every dynamical avenue for vacuum selection that might be available.   The present indications are that if one phenomenologically acceptable vacuum of string theory is found, many others are likely to lie nearby, making vacuum selection on anything other than anthropic grounds impossible unless tunneling between the relevant vacua occurs~\cite{bousso,ashokdouglas,anthropic,douglasdenef,giryavets}.    In the absence of supersymmetry this might yet be possible.

\vspace{0.25in}
{\leftline {\bf Acknowledgements}}
The authors would like to thank: Bobby Acharya, Tom Banks, Volker Braun, Shanta de Alwis, Michael Douglas, Tristan H\"ubsch, Sheldon Katz, Shamit Kachru, Albrecht Klemm, Peter Mayr, Asad Naqvi, Brent Nelson, Sav Sethi.
PB  thanks the Theory Group/Math-Physics Group at the University of Pennsylvania, the Department of Mathematics at University of Durham and the particle theory group at University of California at San Diego for their hospitality at various stages of this project. VB and PB would like to thank the Aspen Center for Physics where this project was initiated, and Addie Berglund for so patiently waiting to be born after this paper was completed.   Work on this project at Penn was supported by the DOE under cooperative research agreement DE-FG02-95ER40893, by the NSF under grant PHY-0331728 and by an NSF Focused Research Grant DMS0139799 for ``The Geometry of Superstrings".   Work on this project at UNH was supported by NSF grant PHY-0355074 and by funds from the College of Engineering and Physical Sciences and the Graduate School at UNH.

\vspace{0.25in}



\end{document}